\documentclass{article-hermes}
\usepackage[dvips]{graphicx}

\begin{document}

\proceedings{20$^{e}$ Journées Bases de Données Avancées (BDA 2004)}{-}

\title[Conception d'un banc d'essais décisionnel]{Conception d'un banc d'essais décisionnel}

\author{Jérôme Darmont \andauthor Fadila Bentayeb \andauthor Omar Boussaïd}

\address{ERIC -- Université Lumière Lyon 2\\
5 avenue Pierre Mendès-France\\
69676 Bron Cedex\\
\small{Contact : \texttt{jerome.darmont@univ-lyon2.fr}}}

\resume{Nous présentons dans cet article un nouveau banc d'essais pour l'évaluation des performances des entrepôts de données. L'emploi de bancs d'essais est profitable pour les utilisateurs, qui peuvent comparer les performances de plusieurs systèmes avant d'investir, mais aussi pour les concepteurs d'entrepôts de données, afin d'évaluer l'impact de différents choix techniques (indexation, matérialisation de vues...). Les bancs d'essais standards édités par le TPC (\emph{Transaction Processing Performance Council}) répondent au premier de ces besoins, mais ne sont pas suffisamment adaptables pour satisfaire le second. C'est pourquoi nous proposons le banc d'essais DWEB (\emph{Data Warehouse Engineering Benchmark}), qui permet de générer à la demande divers entrepôts de données synthétiques, ainsi que les charges (ensembles de requêtes décisionnelles) associées. DWEB est totalement adaptable, mais il demeure facile à mettre en {\oe}uvre grâce à deux niveaux de paramétrage. Par ailleurs, comme DWEB répond principalement à des besoins d'évaluation de performance pour l'ingénierie, il est complémentaire plutôt que concurrent aux bancs d'essais standards du TPC. Finalement, DWEB est implémenté sous la forme d'un logiciel libre écrit en Java qui peut s'interfacer avec la plupart des systèmes de gestion de bases de données relationnels existants.}

\abstract{We present in this paper a new benchmark for evaluating the performances of data warehouses. Benchmarking is useful either to system users for comparing the performances of different systems, or to system engineers for testing the effect of various design choices. While the TPC (\emph{Transaction Processing Performance Council}) standard benchmarks address the first point, they are not tuneable enough to address the second one. Our Data Warehouse Engineering Benchmark (DWEB) allows to generate various ad-hoc synthetic data warehouses and workloads. DWEB is fully parameterized. However, two levels of parameterization keep it easy to tune. Since DWEB mainly meets engineering benchmarking needs, it is complimentary to the TPC standard benchmarks, and not a competitor. Finally, DWEB is implemented as a Java free software that can be interfaced with most existing relational database management systems.}

\motscles{Entrepôts de données, requêtes décisionnelles, OLAP, bancs d'essais, évaluation de performance, conception d'entrepôts de données.}


\maketitlepage

\section{Introduction}
\label{sec:introduction}

Evaluer des technologies centrées sur la décision telles que les entrepôts de données n'est pas une tâche facile. Bien que des conseils pertinents d'ordre général soient disponibles en ligne~\cite{PEN03,GRE04a}, les éléments plus quantitatifs au regard des performances de ces systèmes sont rares.

Dans le contexte des bases de données transactionnelles, des bancs d'essais sont traditionnellement utilisés pour évaluer la performance. Ce sont des modèles synthétiques de bases de données et de charges (opérations à effectuer sur la base), ainsi que des ensembles de mesures de performance. Dans le contexte de l'aide à la décision et plus précisément lors de la conception et de l'exploitation d'un entrepôt de données, de bonnes performances sont encore plus critiques en raison de la nature du modèle de données spécifique et de la volumétrie de ces données. L'objectif de cet article est de proposer un nouveau banc d'essais pour entrepôts de données. 

Plusieurs objectifs peuvent être visés lors de l'utilisation d'un banc d'essais :
\begin{enumerate}
	\item comparer les performances de plusieurs systèmes dans des conditions expérimentales données (utilisateurs de ces systèmes) ;
	\item évaluer l'impact de différents choix architecturaux ou de techniques d'optimisation sur les performances d'un système donné (concepteurs d'entrepôts de données).
\end{enumerate}
Les bancs d'essais proposés par le TPC (\emph{Transaction Processing Performance Council}), un organisme à but non lucratif qui définit des bancs d'essais standards et publie les résultats d'évaluations de performance de façon indépendante, répondent parfaitement au premier de ces objectifs. Cependant, ils ne sont pas très adaptables : leur seul paramètre est un facteur qui définit la taille globale de leur base de données. Néanmoins, dans un contexte de développement, il est intéressant de pouvoir tester une solution (une stratégie d'indexation automatique, par exemple) sur différentes configurations de base de données. Notre objectif est donc de proposer un banc d'essais permettant de générer des entrepôts de données synthétiques, ainsi que les charges (ensembles de requêtes décisionnelles) associées, pour satisfaire des besoins d'ingénierie. Nous l'avons baptisé DWEB (\emph{Data Warehouse Engineering Benchmark}). Il faut souligner que DWEB n'est pas concurrent des bancs d'essais décisionnels standards édités par le TPC. En effet, nous le considérons plutôt comme complémentaire, puisqu'il cible principalement le second objectif mentionné plus haut.

Cet article est organisé comme suit. Nous étudions tout d'abord l'état de l'art concernant les bancs d'essais décisionnels dans la section~\ref{sec:stateoftheart}. Nous détaillons ensuite la base de données et la charge de DWEB dans les sections~\ref{sec:dweb-database} et \ref{sec:dweb-workload}, respectivement. Nous présentons également brièvement notre implémentation de DWEB dans la section~\ref{sec:dweb-implementation}. Nous concluons finalement et présentons nos perspectives de recherche dans la section~\ref{sec:conclusion}.

\section{Bancs d'essais décisionnels existants}
\label{sec:stateoftheart}

À notre connaissance, il existe très peu de bancs d'essais décisionnels en-dehors de ceux du TPC. De plus, les spécifications de ceux que nous avons recensés sont rarement publiées dans leur intégralité~\cite{DEM95}. C'est pourquoi nous nous concentrons sur les bancs d'essais du TPC dans cette section.

\begin{figure*}[hbt]
	\centering
	\includegraphics[width=12cm]{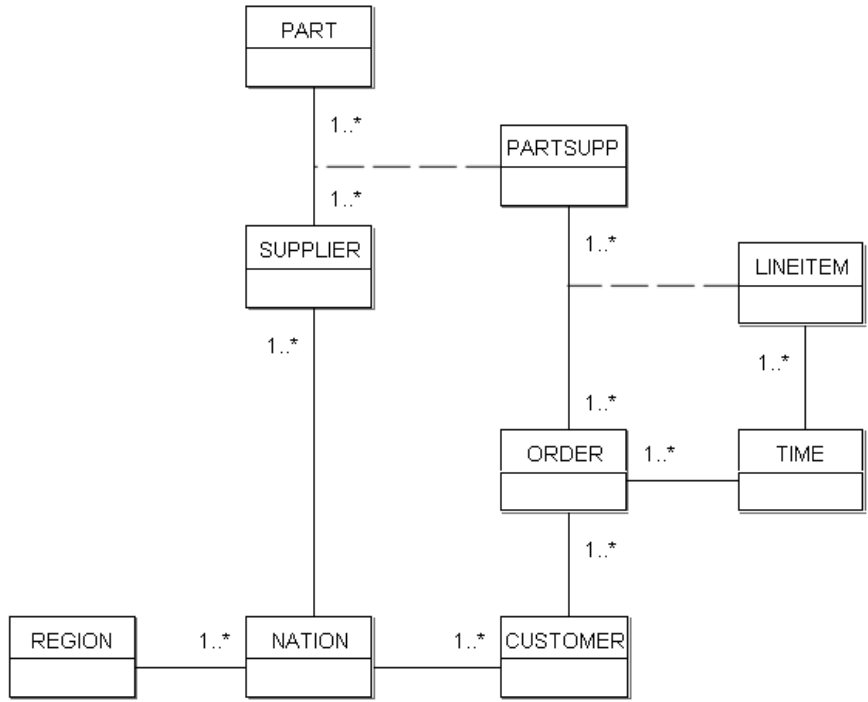}
	\caption{Schéma de la base de données de TPC-D, TPC-H et TPC-R}
	\label{fig:tpc-d}
\end{figure*}

TPC-D~\cite{BAL93,BHA96b,TPC98} a fait son apparition dans le milieu des années 90 et forme la base de TPC-H et TPC-R~\cite{POE00,TPC03a,TPC03b}, qui l'ont désormais remplacé. TPC-H et TPC-R sont en fait identiques, seul leur mode d'utilisation les différencie. TPC-H est conçu pour le requêtage ad-hoc (les requêtes ne sont pas connues à l'avance et toute optimisation préalable est interdite), tandis que TPC-R a une vocation de \emph{reporting} (les requêtes sont connues à l'avance et des optimisations adéquates sont possibles). TPC-H et TPC-R exploitent le même schéma de base de données que TPC-D : un modèle \emph{produits-commandes-fournisseurs} classique (représenté par un diagramme de classes UML dans la figure~\ref{fig:tpc-d}) ; ainsi que la charge de TPC-D enrichie de cinq nouvelles requêtes.

Plus précisément, cette charge est constituée de vingt-deux requêtes décisionnelles paramétrées écrites en SQL-92 et numérotées Q1 à Q22 et de deux fonctions de rafraîchissement RF1 et RF2 qui ajoutent et suppriment des n-uplets dans les tables ORDER et LINEITEM. Les paramètres des requêtes sont instanciés aléatoirement en suivant une loi uniforme. Finalement, le protocole d'exécution de TPC-H ou TPC-R est le suivant :
\begin{enumerate}
	\item un test de chargement ;
	\item un test de performance (exécuté deux fois), lui même subdivisé en un test de puissance et un test de débit.
\end{enumerate}
Trois mesures principales permettent de décrire les résultats obtenus en termes de puissance, de débit et d'une composition de ces deux critères.

\begin{figure*}[hbt]
	\centering
	\includegraphics[width=12cm]{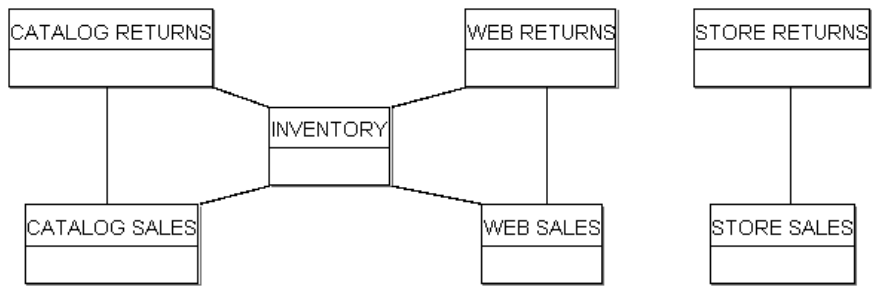}
	\caption{Schéma de l'entrepôt de données de TPC-DS (tables de faits)}
	\label{fig:tpc-ds}
\end{figure*}

TPC-DS, qui est actuellement en cours de développement~\cite{POE02}, modélise plus clairement un entrepôt de données. Il est le successeur annoncé de TPC-H et TPC-R. Le schéma de la base de données de TPC-DS, dont les tables de faits sont représentées dans la figure~\ref{fig:tpc-ds}, représente les fonctions décisionnelles d'un détaillant sous la forme de plusieurs schémas en flocon de neige. Ce modèle comprend également quinze dimensions partagées par les tables de faits. Il s'agit donc d'un schéma en constellation.

La charge de TPC-DS est constituée de quatre classes de requêtes :
requêtes de \emph{reporting},
requêtes décisionnelles ad-hoc,
requêtes interactives d'analyse en ligne (OLAP -- \emph{On-Line Analytical Processing}),
requêtes d'extraction.
Des modèles de requêtes écrits en SQL-99 (et comprenant donc des extensions OLAP) permettent de générer un ensemble d'environ cinq cents requêtes. Ces modèles sont instanciés aléatoirement selon des distributions non-uniformes. Le processus de maintenance de l'entrepôt de données comprend une phase d'ETL (\emph{Extract, Transform, Load}) complète et un traitement spécifique des dimensions. Par exemple, les dimensions historisées conservent les anciens n-uplets quand de nouveaux sont ajoutés, tandis que les dimensions non-historisées ne conservent pas les anciennes données. Finalement, le modèle d'exécution de TPC-DS est divisé en quatre étapes :
\begin{enumerate}
	\item un test de chargement,
	\item une exécution des requêtes,
	\item une phase de maintenance des données,
	\item une seconde exécution des requêtes.
\end{enumerate}
Une seule mesure (de débit) est proposée. Elle prend en compte l'exécution des requêtes et la phase de maintenance.

Bien que les bancs d'essais décisionnels du TPC soient adaptables selon la définition de Gray~\cite{GRA93}, leur schéma est fixe et ils ne sont pas idéaux pour évaluer l'impact de choix architecturaux ou de techniques d'optimisation sur les performances globales. En effet, un seul paramètre permet de définir leur base de données (\emph{Scale Factor} -- $SF$) en terme de taille (de 1 à 100~000~Go). De plus, leur charge n'est pas du tout paramétrable : le nombre de requêtes générées dépend directement de $SF$ dans TPC-DS, par exemple. Il s'avère donc intéressant de proposer un banc d'essais complémentaire capable de modéliser diverses configurations d'entrepôts de données.

\section{Base de données de DWEB}
\label{sec:dweb-database}

\subsection{Schéma}

Notre objectif avec DWEB est de pouvoir modéliser différents types d'architectures d'entrepôts de données~\cite{INM02,KIM02} au sein d'un environnement ROLAP (\emph{Relational OLAP}) :
\begin{itemize}
	\item des schémas en étoile classiques,
	\item des schémas en flocon de neige (avec des dimensions hiérarchisées),
	\item des schémas en constellation (avec plusieurs tables de faits et des dimensions partagées).
\end{itemize}

\begin{figure*}[hbt]
	\centering
	\includegraphics[width=12cm]{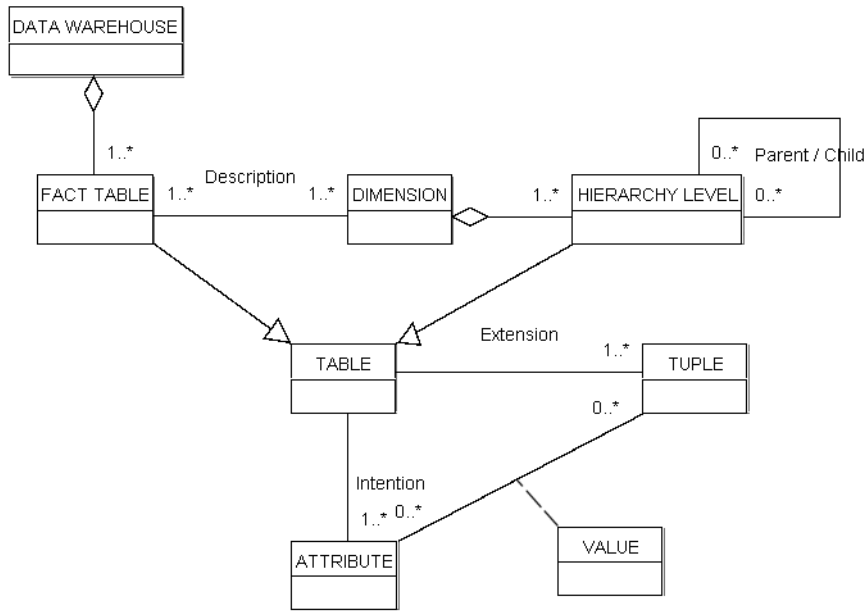}
	\caption{Métaschéma de l'entrepôt de données de DWEB}
	\label{fig:metamodel}
\end{figure*}

Afin d'atteindre ce but, nous proposons un métamodèle d'entrepôt de données (représenté par un diagramme de classes UML dans la figure~\ref{fig:metamodel}) susceptible d'être instancié en ces différents schémas. Nous considérons ce métamodèle comme un intermédiaire entre le métamodèle multidimensionnel du standard CWM (\emph{Common Warehouse Metamodel}) \cite{OMG03,POO03} et le modèle final du banc d'essais. Notre métamodèle est en fait une instance de celui de CWM, qui pourrait en fait être qualifié de méta-métamodèle dans notre contexte.

Notre métamodèle est relativement simple. La partie supérieure de la figure~\ref{fig:metamodel} décrit un entrepôt de données (ou un magasin de données si ce dernier est perçu comme un petit entrepôt dédié) constitué d'une ou plusieurs tables de faits qui sont chacune décrites par plusieurs dimensions. Chaque dimension peut également décrire plusieurs tables de faits (dimensions partagées) et peut contenir une ou plusieurs hiérarchies composées de différents niveaux. Il est possible de n'avoir qu'un seul niveau, auquel cas la dimension n'est pas hiérarchisée.

Les tables de faits et les niveaux hiérarchiques des dimensions sont tous des tables relationnelles, qui sont modélisées dans la partie inférieure de la figure~\ref{fig:metamodel}. Classiquement, une table ou relation est définie en intention par ses attributs et en extension par ses n-uplets. À l'intersection d'un attribut et d'un n-uplet donnés se trouve la valeur de cet attribut dans ce n-uplet.

\subsection{Paramétrage}
\label{db-parameters}

La difficulté principale dans le processus de création d'un schéma d'entrepôt de données est le paramétrage de l'instanciation du métaschéma. Nous avons en effet pour objectif de satisfaire les quatre critères qui définissent, d'après Gray, un \guillemotleft{}~bon~\guillemotright{} banc d'essais~\cite{GRA93} :
\begin{itemize}
	\item \emph{pertinence :} le banc d'essais doit répondre aux besoins d'ingénierie exprimés dans la section~\ref{sec:introduction} ;
	\item \emph{portabilité :} le banc d'essais doit être facile à implémenter sur différents systèmes ;
	\item \emph{adaptabilité :} il doit être possible d'étudier des bases de données de tailles diverses et d'augmenter l'échelle du banc d'essais ;
	\item \emph{simplicité :} le banc d'essais doit être compréhensible sous peine de ne pas être crédible et de demeurer inutilisé.
\end{itemize}

La pertinence et la simplicité forment clairement des objectifs orthogonaux. Introduire des paramètres en trop petit nombre réduit l'expressivité du modèle, tandis qu'en prévoir trop rend le banc d'essais difficile à appréhender par des utilisateurs potentiels. En parallèle, la complexité de génération du schéma instancié doit être maîtrisée. Pour résoudre ce dilemne, nous tirons partie de l'expérience de la conception du banc d'essais orienté objets OCB~\cite{DAR00}. OCB est générique et capable de simuler tous les autres bancs d'essais orientés objets, mais sa mise en {\oe}uvre est contrôlée par des paramètres trop nombreux, si bien que seule une minorité d'entre eux est utilisée en pratique. Nous proposons donc de subdiviser les paramètres en deux sous-ensembles :
\begin{itemize}
	\item un sous-ensemble détaillé de paramètres de bas niveau qui permet à un utilisateur avancé de controler totalement la génération de l'entrepôt de données (tableau~\ref{tab:LowLevelParameters}) ;
	\item une couche supérieure contenant beaucoup moins de paramètres facilement compréhensibles et instanciables (tableau~\ref{tab:HighLevelParameters}). Ces paramètres de haut niveau sont plus précisément les valeurs moyennes des paramètres de bas niveau. Lors de la génération de la base de données, ils sont exploités par des fonctions aléatoires (qui suivent une distribution gaussienne) pour fixer la valeur des paramètres de bas niveau. Finalement, bien que le nombre de paramètres de bas niveau puisse augmenter de façon significative si le schéma grossit, le nombre de paramètres de haut niveau, lui, demeure constant et raisonnable (moins de dix paramètres).
\end{itemize}
Les utilisateurs peuvent alors choisir d'initialiser l'ensemble complet de paramètres de bas niveau ou bien seulement les paramètres de haut niveau, pour lesquels nous proposons des valeurs par défaut correspondant à un schéma en flocon de neige.

\begin{table*}[hbt]
	\centering
		\begin{tabular}{|l|l|}
			\hline 
			{\small \textbf{Nom du paramètre}} & {\small \textbf{Signification}} \\
			\hline \hline
			{\small $NB\_FT$} & {\small Nombre de tables de faits} \\
			\hline
			{\small $NB\_DIM(f)$} & {\small Nombre de dimensions décrivant la table de faits $f$} \\
			\hline
			{\small $TOT\_NB\_DIM$} & {\small Nombre total de dimensions} \\
			\hline
			{\small $NB\_MEAS(f)$} & {\small Nombre de mesures dans la table de faits $f$} \\
			\hline
			{\small $DENSITY(f)$} & {\small Densité de la table de faits $f$} \\
			\hline
			{\small $NB\_LEVELS(d)$} & {\small Nombre de niveaux hiérarchiques dans la dimension $d$} \\
			\hline
			{\small $NB\_ATT(d,h)$} & {\small Nombre d'attributs dans le niveau hiérarchique $h$} \\
			& {\small de la dimension $d$} \\
			\hline
			{\small $HHLEVEL\_SIZE(d)$} & {\small Nombre de n-uplets dans le plus haut niveau hiérarchique} \\
			& {\small de la dimension $d$} \\
			\hline
			{\small $DIM\_SFACTOR(d)$} & {\small Facteur d'échelle pour la taille des niveaux hiérarchiques} \\
			& {\small de la dimension $d$} \\			
			\hline			
		\end{tabular}
	\caption{Paramètres de bas niveau de l'entrepôt de données de DWEB}
	\label{tab:LowLevelParameters}
\end{table*}

\begin{table*}[hbt]
	\centering
		\begin{tabular}{|l|l|c|}
			\hline 
			{\small \textbf{Nom du paramètre}} & {\small \textbf{Signification}} & {\small \textbf{Valeur déf.}} \\
			\hline \hline
			{\small $AVG\_NB\_FT$} & {\small Nombre moyen de tables de faits} & {\small 1} \\
			\hline
			{\small $AVG\_NB\_DIM$} & {\small Nombre moyen de dimensions} & {\small 5} \\
			& {\small par table de fait} & \\
			\hline
			{\small $AVG\_TOT\_NB\_DIM$} & {\small Nombre moyen total de dimensions} & {\small 5} \\
			\hline
			{\small $AVG\_NB\_MEAS$} & {\small Nombre moyen de mesures} & {\small 5} \\
			& {\small dans les tables de faits} & \\
			\hline
			{\small $AVG\_DENSITY$} & {\small Densité moyenne des tables de faits} & {\small 0,6} \\
			\hline
			{\small $AVG\_NB\_LEVELS$} & {\small Nombre moyen de niveaux hiérarchiques} & {\small 3} \\
			&  {\small dans les dimensions} & \\
			\hline
			{\small $AVG\_NB\_ATT$} & {\small Nombre moyen d'attributs} & {\small 5} \\
			& {\small dans les niveaux hiérarchiques} & \\
			\hline
			{\small $AVG\_HHLEVEL\_SIZE$} & {\small Nombre moyen de n-uplets} & {\small 10} \\
			& {\small dans les plus hauts niveaux hiérarchiques} & \\
			\hline
			{\small $DIM\_SFACTOR$} & {\small Facteur d'échelle de taille moyen} & {\small 10} \\
			& {\small au sein des niveaux hiérarchiques} & \\
			\hline			
		\end{tabular}
	\caption{Paramètres de haut niveau de l'entrepôt de données de DWEB}
	\label{tab:HighLevelParameters}
\end{table*}

\textbf{Remarques :}
\begin{itemize}
	\item Puisque des dimensions partagées sont possibles, \\ $TOT\_NB\_DIM \leq \sum{^{NB\_FT}_{i=1}NB\_DIM(i)}$.
	\item La cardinalité d'une table de faits est habituellement inférieure ou égale au produit des cardinalités de ses dimensions. C'est pourquoi nous introduisons la notion de densité. Une densité de 1 indique que toutes les combinaisons possibles des clés primaires des dimensions sont présentes dans la table de faits. Quand la densité diminue, nous éliminons progressivement certaines de ces combinaisons (voir section~\ref{sec:genalgorithm}).
	\item Au sein d'une dimension, un niveau hiérarchique donné a normalement une cardinalité plus grande que le niveau suivant. Par exemple, dans une hiérachie de type \emph{villes-régions-pays}, le nombre de villes doit être plus grand que le nombre de régions, qui doit à son tour être supérieur au nombre de pays. De plus, il existe souvent un facteur d'échelle significatif entre ces cardinalités (par exemple, mille villes, cent régions, dix pays). C'est pourquoi nous définissons la cardinalité des niveaux hiérarchiques en assignant une cardinalité \guillemotleft{}~de départ~\guillemotright{} au plus haut niveau de la hiérarchie ($HHLEVEL\_SIZE$). Nous la multiplions ensuite par un facteur d'échelle prédéfini ($DIM\_SFACTOR$) pour chaque niveau hiérarchique inférieur.
\end{itemize}

\subsection{Algorithme de génération}
\label{sec:genalgorithm}

L'instanciation du métaschéma de DWEB en un schéma d'entrepôt de données réel s'effectue en deux étapes :
\begin{enumerate}
	\item construction des dimensions et de leurs hiérarchies ;
	\item construction des tables de faits.
\end{enumerate}
L'algorithme correspondant à ces deux étapes est présenté dans les figures~\ref{fig:algo-dimensions} et~\ref{fig:algo-fact-tables}, respectivement. Chacune d'entre elles est constituée, pour chaque dimension ou table de fait, de la génération de son intention, puis de son extension. De plus, les hiérarchies des dimensions doivent être gérées. Il faut noter qu'elles sont générées en démarrant du plus haut niveau hiérarchique. Par exemple, pour notre hiérarchie \emph{villes-régions-pays}, nous construisons le niveau \emph{pays} en premier, puis le niveau \emph{région} et enfin le niveau \emph{ville}. Ainsi, les n-uplets d'un niveau hiérarchique donné peuvent référencer ceux du niveau supérieur (qui sont déjà créés) à l'aide d'une clé étrangère.

\begin{figure*}[hbt]
	\centering
	\begin{tabular}{|p{11.5cm}|}
	\hline
	\small{{\tt
	\underline{For} i = 1 \underline{to} TOT\_NB\_DIM \underline{do} \newline
		\hspace*{0.5cm} previous\_ptr = NIL \newline
		\hspace*{0.5cm} size = HHLEVEL\_SIZE(i) \newline
		\hspace*{0.5cm} \underline{For} j = 1 \underline{to} NB\_LEVELS(i) \underline{do} \newline
			\hspace*{1cm} // Intention \newline
			\hspace*{1cm} hl = New(Hierarchy\_level) \newline
			\hspace*{1cm} hl.intention = Primary\_key() \newline
			\hspace*{1cm} \underline{For} k = 1 \underline{to} NB\_ATT(i,j) \underline{do} \newline
				\hspace*{1.5cm} hl.intention = hl.intention $\cup$ String\_descriptor() \newline
			\hspace*{1cm} \underline{End for} \newline
			\hspace*{1cm} // Gestion des hiérarchies \newline
			\hspace*{1cm} hl.child = previous\_ptr \newline
			\hspace*{1cm} hl.parent = NIL \newline
			\hspace*{1cm} \underline{If} previous\_ptr $\neq$ NIL \underline{then} \newline
				\hspace*{1.5cm} previous\_ptr.parent = hl \newline
				\hspace*{1.5cm} hl.intention = hl.intention \newline
				\hspace*{2cm} $\cup$ previous\_ptr.intention.primary\_key // Clé étrangère \newline				
			\hspace*{1cm} \underline{End if} \newline		
			\hspace*{1cm} // Extension \newline
			\hspace*{1cm} hl.extension = $\oslash$ \newline
			\hspace*{1cm} \underline{For} k = 1 \underline{to} size \underline{do} \newline
				\hspace*{1.5cm} new\_tuple = Integer\_primary\_key() \newline
				\hspace*{1.5cm} \underline{For} l = 1 \underline{to} NB\_ATT(i,j) \underline{do} \newline
					\hspace*{2cm} new\_tuple = new\_tuple $\cup$ Random\_string() \newline
				\hspace*{1.5cm} \underline{End for} \newline
				\hspace*{1.5cm} \underline{If} previous\_ptr $\neq$ NIL \underline{then} \newline
					\hspace*{2cm} new\_tuple = new\_tuple $\cup$ Random\_key(previous\_ptr) \newline	
				\hspace*{1.5cm} \underline{End if} \newline	
				\hspace*{1.5cm} hl.extension = hl.extension $\cup$ new\_tuple \newline
			\hspace*{1cm} \underline{End for} \newline							
			\hspace*{1cm} previous\_ptr = hl \newline						
			\hspace*{1cm} size = size * DIM\_SFACTOR(i) \newline	
		\hspace*{0.5cm} \underline{End for} \newline
		\hspace*{0.5cm} dim(i) = hl // Premier (plus bas) niveau de la hiérarchie \newline
	\underline{End for}
	}}\\
	\hline
	\end{tabular}
	\caption{Algotithme de génération des dimensions}
	\label{fig:algo-dimensions}
\end{figure*}

\begin{figure*}[hbt]
	\centering
	\begin{tabular}{|p{11.5cm}|}
	\hline
	\small{{\tt
	\underline{For} i = 1 \underline{to} NB\_FT \underline{do} \newline
		\hspace*{0.5cm} // Intention \newline
		\hspace*{0.5cm} ft(i).intention = $\oslash$ \newline
		\hspace*{0.5cm} \underline{For} j = 1 \underline{to} NB\_DIM(i) \underline{do} \newline		
			\hspace*{1cm} j = Random\_dimension(ft(i)) \newline
			\hspace*{1cm} ft(i).intention = ft(i).intention $\cup$ ft(i).dim(j).primary\_key \newline
		\hspace*{0.5cm} \underline{End for} \newline			
		\hspace*{0.5cm} \underline{For} j = 1 \underline{to} NB\_MEAS(i) \underline{do} \newline		
			\hspace*{1cm} ft(i).intention = ft(i).intention $\cup$ Float\_measure() \newline		
		\hspace*{0.5cm} \underline{End for} \newline	
		\hspace*{0.5cm} // Extension \newline	
		\hspace*{0.5cm} ft(i).extension = $\oslash$ \newline	
		\hspace*{0.5cm} \underline{For} j = 1 \underline{to} NB\_DIM(i) \underline{do}  // Produit cartésien \newline
			\hspace*{1cm} ft(i).extension = ft(i).extension $\times$ ft(i).dim(j).primary\_key\newline	
		\hspace*{0.5cm} \underline{End for} \newline
		\hspace*{0.5cm} to\_delete = DENSITY(i) * |ft(i).extension| \newline
		\hspace*{0.5cm} \underline{For} j = 1 \underline{to} to\_delete \underline{do} \newline	
			\hspace*{1cm} 	Random\_delete(ft(i).extension) \newline		
		\hspace*{0.5cm} \underline{End for} \newline
		\hspace*{0.5cm} \underline{For} j = 1 \underline{to} |ft(i).extension| \underline{do} \newline
		\hspace*{0.5cm} // |ft(i).extension| est supposée à jour \newline			
			\hspace*{1cm} \underline{For} k = 1 \underline{to} NB\_MEAS(i) \underline{do} \newline
				\hspace*{1.5cm} ft(i).extension.tuple(j).measure(k) = Random\_float() \newline
			\hspace*{1cm} \underline{End for} \newline			
		\hspace*{0.5cm} \underline{End for} \newline
	\underline{End for}
	}}\\
	\hline
	\end{tabular}
	\caption{Algorithme de génération des tables de faits}
	\label{fig:algo-fact-tables}
\end{figure*}

Nous utilisons trois classes principales de fonctions et une procédure dans ces algorithmes.
\begin{enumerate}
	\item \texttt{Primary\_key()}, \texttt{String\_descriptor()} et \texttt{Float\_measure()} renvoient des noms pour les clés primaires, les descripteurs  des niveaux hiérarchiques et les mesures des tables de faits, respectivement. Ces noms sont étiquetés séquentiellement et préfixés par le nom de la table (par exemple, DIM1\_1\_DESCR1, DIM1\_1\_DESCR2...).
	\item \texttt{Integer\_primary\_key()}, \texttt{Random\_key()}, \texttt{Random\_string()} et \texttt{Random\_float()} renvoient des entiers séquentiels par rapport à une table donnée (aucun doublon n'est permis), des instances aléatoires de la clé primaire de la table spécifiée (valeurs aléatoires pour une clé étrangère), des chaînes de caractères aléatoires de taille fixe (20 caractères) sélectionnées dans un référentiel préalablement construit et préfixées par le nom de l'attribut correspondant, ainsi que des nombres aléatoires réels simple précision, respectivement.
	\item \texttt{Random\_dimension()} renvoie une dimension sélectionnée parmi les dimensions existantes qui ne décrivent pas déjà la table de faits passée en argument.
	\item \texttt{Random\_delete()} efface aléatoirement un n-uplet dans l'extension d'une table.
\end{enumerate}
A l'exception de la procédure \texttt{Random\_delete()}, dans laquelle nous utilisons une distribution aléatoire uniforme, nous employons des distributions aléatoires gaussiennes.

\textbf{Remarque :} La manière dont la densité est gérée dans la figure~\ref{fig:algo-fact-tables} est clairement non-optimale. Nous choisissons de présenter l'algorithme de cette manière afin de le rendre plus clair, mais notre implémentation ne crée pas tous les n-uplets du résultat du produit cartésien avant d'en effacer certains. Nous générons directement le bon nombre de n-uplets en utilisant la densité comme une probabilité de création de chaque n-uplet.

\section{Charge de DWEB}
\label{sec:dweb-workload}

Dans un banc d'essais pour entrepôts de données, la charge peut être subdivisée en :
\begin{itemize}
	\item une charge de requêtes décisionnelles (principalement des requêtes OLAP) ;
	\item le processus d'ETL (chargement et rafraîchissement des données).
\end{itemize}

TPC-DS gère ces deux aspects. Puisque la base de données de DWEB compose la partie la plus originale de notre travail, nous nous inspirons donc de la définition de charge de TPC-DS, tout en exploitant d'autres sources d'informations concernant la performance des entrepôts de données~\cite{BMC00,GRE04b}. Néanmoins, ils nous est nécessaire d'adapter les propositions de TPC-DS à la nature variable des entrepôts de données de DWEB. De plus, nous voulons satisfaire le critère de simplicité de Gray, aussi proposons-nous au final une charge plus simple que celle de TPC-DS.

Par ailleurs, nous nous concentrons dans un premier temps sur la définition d'un modèle de requêtes. La modélisation du processus d'ETL complet est une tâche complexe sur laquelle nous comptons revenir plus tard. Nous considérons pour l'instant que les spécifications actuelles de DWEB permettent une évaluation grossière d'un chargement d'entrepôt. En effet, la base de données de DWEB peut être générée sous forme de fichiers plats, puis ensuite chargée dans un entrepôt en utilisant les outils fournis par le système.

\subsection{Modèle de requêtes}
\label{sec:query-model}

La charge de DWEB modélise deux types de requêtes :
\begin{itemize}
	\item des requêtes purement décisionnelles contenant les opérations OLAP usuelles telles que cube, agrégation (\emph{roll-up}), forage (\emph{drill down}) et projection/sélection (\emph{slice and dice}) ;
	\item des requêtes d'extraction.
\end{itemize}

Nous définissons notre modèle de requêtes générique (figure~\ref{fig:query-model}) à l'aide d'une grammaire qui est un sous-ensemble du standard SQL-99, qui introduit des outils analytiques (indispensables dans un contexte décisionnel) dans les requêtes relationnelles. Cela nous permet de générer des requêtes SQL analytiques dynamiquement.

\begin{figure*}[hbt]
	\centering

\begin{tabular}{|ll|}
\hline
  \small{Query ::-} & \\
  &\\
  \small{{\bf Select}} & \small{![$<$Attribute Clause$>$ $|$ $<$Aggregate Clause$>$} \\&\small{$|$ [$<$Attribute Clause$>$\textbf{,} $<$Aggregate Clause$>$]]} \\
  \small{{\bf From}} & \small{!$<$Table Clause$>$ [$<$Where Clause$>$} \\
   & \small{$\|$ [$<$Group by Clause$>$ *$<$Having Clause$>$]]} \\
  & \\

  \small{Attribute Clause ::-} &  \small{{\it Attribute Name} [[\textbf{,} $<$Attribute Clause$>$] $|$ $\bot$]} \\
  \small{Aggregate Clause ::-} & \small{![{\it Aggregate Function Name} \textbf{(}{\it Attribute Name}\textbf{)}] [{\bf As}  {\it Alias}]}\\
  &  \small{[[\textbf{,} $<$Aggregate Clause$>$] $|$ $\bot$]}\\
  &\\
  & \\
  \small{Table Clause ::-} &  \small{{\it Table Name} [[\textbf{,} $<$Table Clause$>$] $|$$\bot$]}\\
  & \\
  \small{Where Clause ::-}&  \small{{\bf Where} ![$<$Condition Clause$>$ $|$ $<$Join Clause$>$} \\&\small{$|$ [$<$Condition Clause$>$ \textbf{And} $<$Join Clause$>$]]} \\
  \small{Condition Clause ::-} & \small{![{\it Attribute Name} $<$Comparison Operator$>$ $<$Operand Clause$>$]}\\
  & \small{[[$<$Logical Operator$>$ $<$Condition Clause$>$] $|$ $\bot$]}\\
  \small{Operand Clause ::-} &  \small{[{\it Attribute Name} $\mid$ {\it Attribute Value} $\mid$ {\it Attribute Value List}]} \\
 \small{Join Clause ::-} & \small{![{\it Attribute Name} i = {\it Attribute Name} j] [[\textbf{And} $<$Join Clause$>$] $|$$\bot$]} \\
 & \\
 \small{Group by Clause ::-} & \small{{\bf Group by} [{\bf Cube} $\mid$ {\bf Rollup}] $<$Attribute Clause$>$} \\
 \small{Having Clause ::-} & \small{[{\it Alias} $\mid$ {\it Aggregate Function Name} \textbf{(}{\it Attribute Name}\textbf{)}]}\\
 & \small{$<$Comparison Operator$>$  [{\it Attribute Value} $\mid$ {\it Attribute Value List}]}\\
\hline 
\end{tabular}

	\caption{Modèle de requêtes de DWEB}
	\label{fig:query-model}
\end{figure*}

Définissons la sémantique de la terminologie employée dans la figure~\ref{fig:query-model}.
\begin{itemize}
	\item Les crochets [ et ] sont des délimiteurs.
  \item !$<$A$>$ : A est requis.
  \item *$<$A$>$ : A est optionnel.
  \item $<$A $\|$ B$>$ : A ou B.
  \item $<$A $\mid$ B$>$ : A ou exclusif B.
  \item $\bot$ : clause vide.
  \item Les éléments du langage SQL sont indiqués en gras.
\end{itemize}

\subsection{Paramétrage}

Comme les paramètres de la base de données de DWEB (section~\ref{db-parameters}), ceux qui définissent sa charge (tableau~\ref{tab:workload-parameters}) ont été conçus pour répondre au critère de simplicité de Gray. Ils déterminent la manière dont le modèle de requêtes de la figure~\ref{fig:query-model} s'instancie. Nous ne disposons ici que d'un nombre limité de paramètres de haut niveau (huit paramètres, puisque $PROB\_EXTRACT$ et $PROB\_ROLLUP$ sont calculés à partir de $PROB\_OLAP$ et $PROB\_CUBE$, respectivement). Il n'est en effet pas envisageable d'entrer plus dans le détail si la taille de la charge atteint cinq cents requêtes comme c'est le cas dans TPC-DS, par exemple.

\begin{table*}[hbt]
	\centering
		\begin{tabular}{|l|l|c|}
			\hline 
			{\small \textbf{Nom du paramètre}} & {\small \textbf{Signification}} & {\small \textbf{Valeur par défaut}} \\
			\hline \hline
			{\small $NB\_Q$} & {\small Nombre approximatif de requêtes} & {\small 100} \\
			& {\small dans la charge} & \\
			\hline
			{\small $AVG\_NB\_ATT$} & {\small Nombre moyen d'attributs sélectionnés} & {\small 5} \\
			& {\small dans une requête} & \\
			\hline
			{\small $AVG\_NB\_RESTR$} & {\small Nombre moyen de restrictions} & {\small 3} \\
			& {\small dans une requête} & \\
			\hline
			{\small $PROB\_OLAP$} & {\small Probabilité que le type} & {\small 0,9} \\
			& {\small de requête soit OLAP} & \\
			\hline
			{\small $PROB\_EXTRACT$} & {\small Probabilité que la requête} & {\small $1$} \\	
			& {\small soit une requête d'extraction} & {\small $- PROB\_OLAP$}\\
			\hline
			{\small $AVG\_NB\_AGGREG$} & {\small Nombre moyen d'agrégations} & {\small 3} \\
			& {\small dans une requête OLAP} & \\
			\hline
			{\small $PROB\_CUBE$} & {\small Probabilité qu'une requête OLAP} & {\small 0,3} \\
			& {\small utilise l'opérateur $Cube$} & \\
			\hline
			{\small $PROB\_ROLLUP$} & {\small Probabilité qu'une requête OLAP} & {\small $1$} \\
			& {\small utilise l'opérateur $Rollup$} & {\small $- PROB\_CUBE$} \\
			\hline
			{\small $PROB\_HAVING$} & {\small Probabilité qu'une requête OLAP} & {\small 0,2} \\
			& {\small contienne une clause $Having$} & \\
			\hline			
			{\small $AVG\_NB\_DD$} & {\small Nombre moyen de forages (\emph{drill downs})} & {\small 3} \\
			& {\small après une requête OLAP} & \\
			\hline
		\end{tabular}
	\caption{Paramètres de la charge de DWEB}
	\label{tab:workload-parameters}
\end{table*}

\textbf{Remarque :} $NB\_Q$ n'est qu'une approximation du nombre de requêtes parce que le nombre de forages (\emph{drill downs}) effectués après une requête OLAP est variable. Nous ne pouvons donc arrêter le processus de génération que lorsque nous avons effectivement produit au moins $NB\_Q$ requêtes.

\subsection{Algorithme de génération}

L'algorithme de génération de la charge de DWEB est présenté dans les figures~\ref{fig:algo-workload1} et~\ref{fig:algo-workload2}. Son objectif est de générer un ensemble de requêtes SQL-99 qui puisse être directement exécuté sur l'entrepôt de données synthétique défini à la section~\ref{sec:dweb-database}. Il est subdivisé en deux étapes :
\begin{enumerate}
	\item génération d'une requête initiale qui peut être soit une requête OLAP, soit une requête d'extraction ;
	\item si la requête initiale était de type OLAP, exécution d'un certain nombre de forages basés sur cette première requête. Plus précisément, à chaque fois qu'un nouveau forage est exécuté, un attribut d'un niveau inférieur de la hiérarchie d'une dimension est ajouté à la clause $Attribute~Clause$ de la requête précédente.
\end{enumerate}
La première étape est elle-même subdivisée en trois sous-étapes :
\begin{enumerate}
	\item les clauses $Select$, $From$ et $Where$ d'une requête sont générées simultanément en sélectionnant aléatoirement une table de faits et des dimensions (y compris un niveau hiérarchique dans chacune de ces dimensions) ;
	\item la clause $Where$ est complétée par des conditions supplémentaires ;
	\item finalement, le choix du type de requête est effectué (requête OLAP ou requête d'extraction). Dans le second cas, la requête est terminée. Dans le premier, des fonctions d'agrégat appliquées à des mesures de la table de faits sont ajoutées à la requête, ainsi qu'une clause $Group~by$ pouvant inclure les opérateurs $Cube$ ou $Rollup$. Il est également possible d'ajouter une clause $Having$. La fonction d'agrégat que nous appliquons sur les mesures est toujours une somme car c'est l'opération la plus courantes dans les cubes. De plus, les autres fonctions d'agrégat ont des complexités similaires en temps de calcul. Elles n'apporteraient donc pas plus d'information dans le cadre d'une étude de performance.
\end{enumerate}

\begin{figure}[htb]
	\centering
	\begin{tabular}{|p{11.5cm}|}
	\hline
	\small{{\tt
	n = 0 \newline
	\underline{While} n < NB\_Q \underline{do} \newline		
		\hspace*{0.5cm} // Etape 1 : Requête initiale \newline
		\hspace*{0.5cm} // Etape 1.2 : Clauses Select, From et Where \newline
		\hspace*{0.5cm} i = Random\_FT() // Sélection de la table de faits \newline
		\hspace*{0.5cm} attribute\_list = $\oslash$ \newline
		\hspace*{0.5cm} table\_list = ft(i) \newline
		\hspace*{0.5cm} condition\_list = $\oslash$ \newline
		\hspace*{0.5cm} \underline{For} k = 1 \underline{to} Random\_int(AVG\_NB\_ATT) \underline{do} \newline		
			\hspace*{1cm} j = Random\_dimension(ft(i)) // Sélection d'une dimension \newline
			\hspace*{1cm} l = Random\_int(1, ft(i).dim(j).nb\_levels) \newline
			\hspace*{1cm} // Positionnement au niveau hiérarchique l \newline
			\hspace*{1cm} hl = ft(i).dim(j) // Niveau hiérarchique courant \newline
			\hspace*{1cm} m = 1 // Compteur de niveau \newline
			\hspace*{1cm} fk = ft(i).intention.primary\_key.element(j) \newline
			\hspace*{1cm} // (Cette clé étrangère correspond à ft(i).dim(j).primary\_key) \newline
			\hspace*{1cm} \underline{While} m < l \underline{and} hl.child $\neq$ NIL \underline{do} \newline		
				\hspace*{1.5cm} // Construction de la jointure \newline
				\hspace*{1.5cm} table\_list = table\_list $\cup$ hl \newline
				\hspace*{1.5cm} condition\_list = condition\_list \newline
					\hspace*{2cm}$\cup$ (fk = hl.intention.primary\_key) \newline
				\hspace*{1.5cm} // Niveau suivant \newline
				\hspace*{1.5cm} fk = hl.intention.foreign\_key \newline
				\hspace*{1.5cm} m = m + 1 \newline
				\hspace*{1.5cm} hl = hl.child \newline
			\hspace*{1cm} \underline{End while} \newline	
			\hspace*{1cm} attribute\_list = attribute\_list \newline
				\hspace*{1.5cm}$\cup$ Random\_attribute(hl.intention) \newline
		\hspace*{0.5cm} \underline{End for} \newline			
		\hspace*{0.5cm} // Etape 1.2 : Compléter la clause Where \newline			
		\hspace*{0.5cm} \underline{For} k = 1 \underline{to} Random\_int(AVG\_NB\_RESTR) \underline{do} \newline
			\hspace*{1cm} condition\_list = condition\_list \newline
				\hspace*{1.5cm} $\cup$ (Random\_attribute(attribute\_list) = Random\_string()) \newline
		\hspace*{0.5cm} \underline{End for} \newline	
		\hspace*{0.5cm} // Etape 1.3 : Sélection requête OLAP ou requête d'extraction \newline	
		\hspace*{0.5cm} p1 = Random\_float(0,1) \newline	
		\hspace*{0.5cm} \underline{If} p1 $\leq$ PROB\_OLAP \underline{then} // Requête OLAP \newline
			\hspace*{1cm} // Agrégat \newline
			\hspace*{1cm} aggregate\_list = $\oslash$ \newline
						\hspace*{1cm} \underline{For} k = 1 \underline{to} Random\_int(AVG\_NB\_AGGREG) \underline{do} \newline
				\hspace*{1.5cm} aggregate\_list = aggregate\_list \newline 
					\hspace*{2cm} $\cup$ Random\_measure(ft(i).intention) \newline
			\hspace*{1cm} \underline{End for} \newline
					../..
	}}\\
	\hline
	\end{tabular}
	\caption{Algorithme de génération de la charge -- Partie 1}
	\label{fig:algo-workload1}
\end{figure}

\begin{figure}[htb]
	\centering
	\begin{tabular}{|p{11.5cm}|}
	\hline
	\small{{\tt
		../..			\newline
								\hspace*{1cm} // Clause Group by \newline
			\hspace*{1cm} group\_by\_list = attribute\_list \newline
			\hspace*{1cm} p2 = Random\_float(0,1) \newline	
			\hspace*{1cm} \underline{If} p2 $\leq$ PROB\_CUBE \underline{then} \newline
				\hspace*{1.5cm} group\_by\_operator = CUBE \newline	
			\hspace*{1cm} \underline{Else} \newline	
				\hspace*{1.5cm} group\_by\_operator = ROLLUP \newline				
			\hspace*{1cm} \underline{End if} \newline	
					\hspace*{1cm} // Clause Having \newline
			\hspace*{1cm} p3 = Random\_float(0,1) \newline	
			\hspace*{1cm} \underline{If} p3 $\leq$ PROB\_HAVING \underline{then} \newline
				\hspace*{1.5cm} having\_clause = (Random\_attribute(aggregate\_list), $\geq$, \newline
					\hspace*{2cm} Random\_float()) \newline	
			\hspace*{1cm} \underline{Else} \newline	
				\hspace*{1.5cm} having\_clause = $\oslash$ \newline				
			\hspace*{1cm} \underline{End if} \newline	
					\hspace*{0.5cm} \underline{Else} // Requête d'extraction \newline		
			\hspace*{1cm} group\_by\_list = $\oslash$ \newline	
			\hspace*{1cm} group\_by\_operator = $\oslash$ \newline	
			\hspace*{1cm} having\_clause = $\oslash$ \newline									
		\hspace*{0.5cm} \underline{End if} \newline
		\hspace*{0.5cm} // Génération de la requête SQL \newline
		\hspace*{0.5cm} Gen\_query(attribute\_list, aggregate\_list, table\_list,  \newline
			\hspace*{1cm} condition\_list, group\_by\_list, group\_by\_operator, \newline
			\hspace*{1cm} having\_clause) \newline
		\hspace*{0.5cm} n = n + 1 \newline			
		\hspace*{0.5cm} // Etape 2 : Eventuelles requêtes DRILL DOWN \newline
		\hspace*{0.5cm} \underline{If} p1 $\leq$ PROB\_OLAP \underline{then} \newline		
			\hspace*{1cm} k = 0 \newline
			\hspace*{1cm} \underline{While} k < Random\_int(AVG\_NB\_DD) \underline{and} hl.parent $\neq$ NIL \underline{do} \newline						
				\hspace*{1.5cm} k = k +1 \newline
				\hspace*{1.5cm} hl = hl.parent \newline
				\hspace*{1.5cm} att = Random\_attribute(hl.intention) \newline
				\hspace*{1.5cm} attribute\_list = attribute\_list $\cup$ att \newline
				\hspace*{1.5cm} group\_by\_list = group\_by\_list $\cup$ att \newline
				\hspace*{1.5cm} Gen\_query(attribute\_list, aggregate\_list, table\_list,  \newline
					\hspace*{2cm} condition\_list, group\_by\_list, group\_by\_operator, \newline
					\hspace*{2cm} having\_clause) \newline				
			\hspace*{1cm} \underline{End while} \newline	
			\hspace*{1cm} n = n + k \newline			
		\hspace*{0.5cm} \underline{End if} \newline						
	\underline{End while}
	}}\\
	\hline
	\end{tabular}
	\caption{Algorithme de génération de la charge -- Partie 2}
	\label{fig:algo-workload2}
\end{figure}

Nous utilisons trois classes de fonctions et une procédure dans cet algorithme.
\begin{enumerate}
	\item \texttt{Random\_string()} et \texttt{Random\_float()} sont les mêmes fonctions que celles décrites dans la section~\ref{sec:genalgorithm}. Cependant, nous introduisons ici la possibilité pour \texttt{Random\_float()} d'utiliser une distribution aléatoire soit uniforme, soit gaussienne. Cela dépend des paramètres passés à la fonction : distribution uniforme pour un intervalle de valeurs, distribution gaussienne pour une valeur moyenne. Finalement, nous introduisons la fonction \texttt{Random\_int()} qui se comporte exactement comme \texttt{Random\_float()}, mais qui retourne des valeurs entières.
	\item \texttt{Random\_FT()} et \texttt{Random\_dimension()} permettent de sélectionner une table de faits et une dimension qui décrit une table de faits donnée, respectivement. Elles utilisent toutes deux une distribution aléatoire gaussienne. \texttt{Random\_dimension()} est également déjà décrite dans la section~\ref{sec:genalgorithm}.
	\item \texttt{Random\_attribute()} et \texttt{Random\_measure()} ont des comportements très similaires. Elles retournent un attribut ou une mesure, respectivement, appartenant à l'intention d'une table ou à une liste d'attributs. Elles utilisent toutes deux une distribution aléatoire gaussienne.
	\item \texttt{Gen\_query()} est la procédure qui est chargée de générer effectivement le code SQL-99 des requêtes de la charge, d'après les paramètres nécessaires pour instancier notre modèle de requêtes.
\end{enumerate}

\section{Implémentation de DWEB}
\label{sec:dweb-implementation}

DWEB est implémenté sous la forme d'une application Java. Nous avons choisi le langage Java pour satisfaire le critère de portabilité de Gray. La version actuelle de notre prototype permet de générer des schémas en flocon de neige. Les schémas en constellation ne sont pas encore implémentés. Par ailleurs, comme les paramètres de DWEB peuvent sembler abstraits, notre prototype propose une estimation de la taille de l'entrepôt généré en méga-octets une fois que les paramètres sont fixés et avant la génération de la base de données. Ainsi, les utilisateurs peuvent ajuster les paramètres pour obtenir le type d'entrepôt de données qu'ils souhaitent. Notre prototype peut être interfacé à la plupart des systèmes de gestion de bases de données relationnels grâce à JDBC. La charge de DWEB est également en cours d'impléméntation. Seule une charge simplifiée est disponible actuellement.

Par ailleurs, comme nous utilisons de nombreuses fonctions aléatoires, nous pensons aussi introduire dans notre prototype un générateur pseudo-aléatoire plus performant que ceux fournis en standard, comme celui proposé par Lewis et Payne~\cite{LEW73}, qui est un des meilleurs grâce à sa très grande période.

Finalement, bien que notre application soit en perpétuelle évolution, sa dernière version est en permanence disponible en ligne~\cite{DWEB04}.

\section{Conclusion et perspectives}
\label{sec:conclusion}

Nous avons présenté dans cet article les spécifications complètes d'un nouveau banc d'essais pour entrepôts de données baptisé DWEB. La caractéristique principale de DWEB est qu'il permet de générer des entrepôts de données synthétiques variés, ainsi que les charges (ensemble de requêtes) associées. Les schémas d'entrepôt classiques, comme les modèles en étoile, en flocon de neige et en constellation sont en effet supportés. Nous considérons DWEB comme un banc d'essais d'ingénierie destiné aux concepteurs d'entrepôts de données et de systèmes. En cela, il n'est pas concurrent du banc d'essais TPC-DS (actuellement en cours de développement), qui est plus à destination des utilisateurs finaux, pour des évaluations de performances \guillemotleft{}~pures~\guillemotright{}. Il faut toutefois remarquer que le schéma d'entrepôt de données de TPC-DS peut être modélisé à l'aide de DWEB. Cependant, la charge de DWEB n'est actuellement pas aussi élaborée que celle de TPC-DS. Finalement, nous avons pour objectif de fournir dans cet article les spécifications les plus complètes possibles de DWEB, de manière à ce que notre banc d'essais puisse être implémenté facilement par d'autres chercheurs et/ou concepteurs d'entrepôts de données.

Nos travaux futurs dans ce domaine sont divisés en quatre axes. Premièrement, il nous est nécessaire de terminer l'implémentation de DWEB (charge, schémas en constellation avec hiérarchies partagées, génération de fichier plats et chargement dans une base de données, tests divers...). Ce travail est actuellement en cours. Des expériences avec DWEB devraient également nous permettre de proposer un meilleur paramétrage par défaut de notre banc d'essais. Nous encourageons également d'autres chercheurs et/ou concepteurs d'entrepôts de données à publier les résultats de leurs propres expériences avec DWEB.

Deuxièmement, ils nous faut tester la pertinence (d'après la définition de Gray) de notre banc d'essais pour l'évaluation de performance des entrepôts de données dans un contexte d'ingénierie. Afin d'atteindre cet objectif, nous pensons comparer l'efficacité de plusieurs techniques de sélection automatique d'index et de vues matérialisées (dont certaines de nos propositions, ce qui a constitué une des motivations de la conception de DWEB à l'origine).

Troisièmement, nous comptons progressivement améliorer DWEB. Par exemple, dans cet article, nous supposons implicitement que nous pouvons réutiliser ou facilement adapter le protocole d'exécution et les mesures de performance de TPC-DS. Une réflexion plus élaborée à propos de ces sujets pourrait être intéressante. Il sera également important d'inclure complètement le processus d'ETL dans notre charge. Plusieurs travaux existants~\cite{LAB98} pourraient d'ailleurs nous y aider. Finalement, il serait intéressant de rendre notre charge plus dynamique, comme cela a été fait dans la plate-forme d'évaluation de performance DoEF~\cite{HE03,isi04}. Bien que la dynamicité ne soit peut-être pas un facteur pertinent dans une charge décisionnelle, cette possibilité pourrait néanmoins être explorée. Pour finir, puisque nous travaillons par ailleurs sur l'entreposage de données complexes (multiformats, multistructures, multisources, multimodales et/ou multiversions), nous envisageons également une extension \guillemotleft{}~données complexes~\guillemotright{} pour DWEB à long terme.

Finalement, DWEB pourrait évoluer pour devenir capable de proposer automatiquement des configurations d'entrepôts de données et des charges afin d'évaluer les performances d'une architecture d'entrepôt donnée ou d'une technique de sélection d'index ou de vue matérialisée. DWEB pourrait alors capitaliser la connaissance acquise au cours de diverses expériences et la réutiliser pour proposer de nouvelle configurations, plus pertinentes.

\bibliography{ds_benchmarks}

\end{document}